# Preliminary design of laser accelerator beam line

Y.Shang, K.Zhu*, C.Cao, J.G.Zhu, Y.R.Lu, Z.Y.Guo, J.E.Chen, X.Q.Yan*

State Key Laboratory of Nuclear Physics and Technology & Key Lab of High Energy Density Physics Simulation, CAPT, Peking University, Beijing 100871, China

**Abstract:** A Compact laser plasma accelerator (CLAPA) is being built in Peking University, which is based on RPA-PSA mechanism or other acceleration mechanisms. According to the beam parameters from preparatory experiments and theoretical simulations, the beam line is preliminarily designed. The beam line is mainly constituted by common transport elements to deliver proton beam with the energy of 1~50MeV, energy spread of 0~±1% and current of 0~$10^8$ proton per pulse to satisfy the requirement of different experiments. The simulation result of 15MeV proton beam with an energy spread of ±1%, current of $1\times10^8$ proton per pulse and final spot radius of 9mm is presented in this paper.

## 1 Introduction

High energy particles can be produced when laser interacts with the plasma [1]. The laser accelerator can acquire $10^{12}$V/m accelerating electric field gradient which is at least 1000 times higher than conventional accelerators [2]. The laser accelerator can accelerate ions more effectively and greatly reduce the scale and cost. It has promising prospects in compact ultra high energy ion accelerator and many applications [3].

A compact laser accelerator (CLAPA), which is according to RPA-PSA mechanism [4-6] or other acceleration mechanisms [7], will be built in Peking University. Many researches will be carried on CLAPA: basic research in physical mechanism of laser-plasma acceleration, ultra short and intense pulse beam transport, self-supporting ultra-thin target production; application research in medicine [8], inertial confinement fusion (ICF) [9], astrophysics etc.

At present, Laser accelerators are studied in many institutes around the world [10], but the application research is still in an exploratory stage [11-19]. The reason is that the beam produced by laser accelerator has the characteristic of short duration, high pulse current and wide energy spectrum [20]. The beam pulse duration is only tens of picoseconds and $10^8$-$10^{10}$ ions are produced in one pulse [21], so the peak current can reach ampere scale. The initial beam spot is as small as laser spot, which means the beam spot radius is a few microns. There is no doubt that the space charge effect will be very strong. Although the beam contains co-moving electrons which can neutralize the space charge effect at the initial, these electrons will be moved out of the beam under the effect of transport elements. So, the initial collection and collimation is a very difficult and critical part of the beam line. Many kinds of elements are tried, like permanent magnet quadrupole lens [12] [13], solenoid magnet [14] [15], laser triggered micro-lens [16]. Particle selection is another critical part. Because the beam has wide energy spectrum, the chromatic aberration of collected beam is serious. The beam must be selected with desired energy spectra. It usually depends on bending magnet [17] or a set of dipole magnets [18] [19].

To solve the problems above, we preliminarily design a beam line for CLAPA. The beam line is mainly constituted by common transport elements to deliver proton beam with the energy of 1~50MeV, energy spread of 0~±1% and current of 0~$10^8$ proton per pulse

## 2 Beam line

In this section, we mainly talk about the beam line on request of biomedical irradiation. The beam parameters are shown in Tab.1. The simulation of the higher energy beam transport, like 50MeV proton beam, is shown in the end.

Tab.1 beam parameters of CLAPA

| | |
|---|---|
| ion | proton |
| energy | 15MeV |
| current | $1\times10^8$proton/pulse |
| initial energy spread | ±15% |
| final energy spread | ±1% |
| initial transverse radius | 0.005mm |
| initial longitudinal length | 1.06mm |
| Final transverse radius | 9mm |
| Final longitudinal length | 70mm |

The beam line is constituted by two parts: collection part and analysis part. In collection part, the aperture is used to remove big divergence angle protons as more as possible at the initial. Then, the beam is collected by a quadrupole-triplet lens and a quadrupole-doublet lens. In analysis part, the beam is analyzed by a bending magnet. Finally, the beam is focused by another quadrupole-doublet lens and delivered to experiment platform. The structure of beam line is shown in Fig.1.

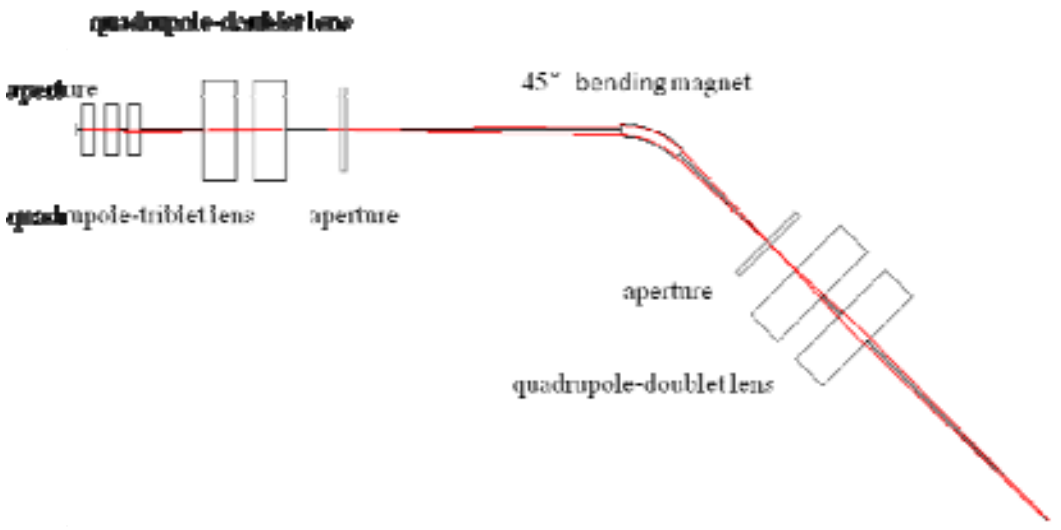

Fig.1 Schematic diagram of beam line

First order simulation of transport is carried by program Trace-3D and high order simulation is carried by program Track [22]. The simulation results are shown in Fig.2.

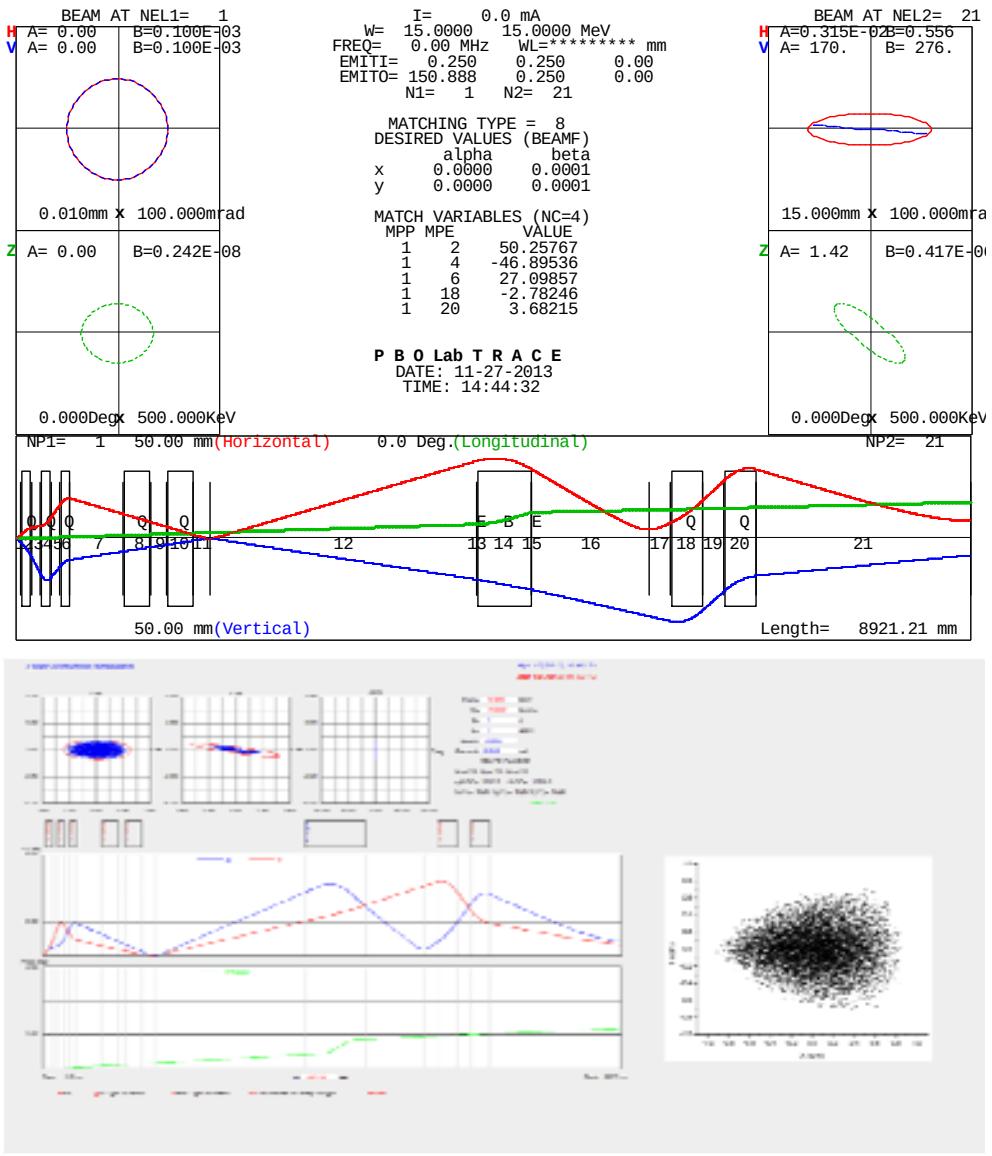

Fig.2 Simulation results of proton beam transport

The upper is the result of Trace-3D and the lower is the result of Track

Comparing the results of two programs, the beam envelopes are almost the same except the size of final spots. The difference is caused by chromatic aberration. That is to say, the big energy spread causes the difference. There is no good way to deal with the chromatic aberration, but reducing the energy spread after analyzing. The detail of beam line structure is presented below.

Aperture: Due to the laser acceleration mechanism and high space charge effect, the beam has big divergence angle. It is impossible to deliver all of the protons to the end. To avoid the influence of big divergence angle ions and reduce the space charge effect, the beam is screened by an aperture at the beginning of transport. The radius of aperture is 3mm. The distance is 50mm away from laser target. The proton beam passes through the aperture with divergence angle of ±50mrad, transverse emittance of 0.25π mm.mrad and current of $1\times10^8$ proton/pulse.

Collecting lens: After passing the aperture,

the beam will expand fast in transverse direction. It is necessary to focus the beam as early as possible to avoid the unnecessary losses. A quadrupole-triplet lens is designed to focus the beam which is adjoined to the aperture. The inner radius of lens is 20mm, the length of lens is 100mm and the distance is 80mm between each other. When the magnet fields of lenses are 5.00, -4.71 and 2.75 kG/cm, the proton beam can be perfectly collected.

Assistant collecting lens: The collecting lens can collect the protons with the energy lower than 30MeV. But it is difficult to collect the protons with higher energy because of the limit of magnet field. So, a quadrupole-doublet lens is added to assist collection. The inner radius of lens is 40mm, the length of lens is 250mm and the distance is 150mm between each other.

Bending magnet: The proton beam produced by laser accelerator has wide energy spectrum and a lot of different ions. Although the different ions may be screened partly in the collecting stage due to the chromatic aberration, the proton beam still contains different ions. To get the high quality proton beam, a 45° bending magnet is used to analyze beam. The radius of bending magnet is 650mm. The analyzing ability of bending magnet is simulated by program Track (Fig.3).

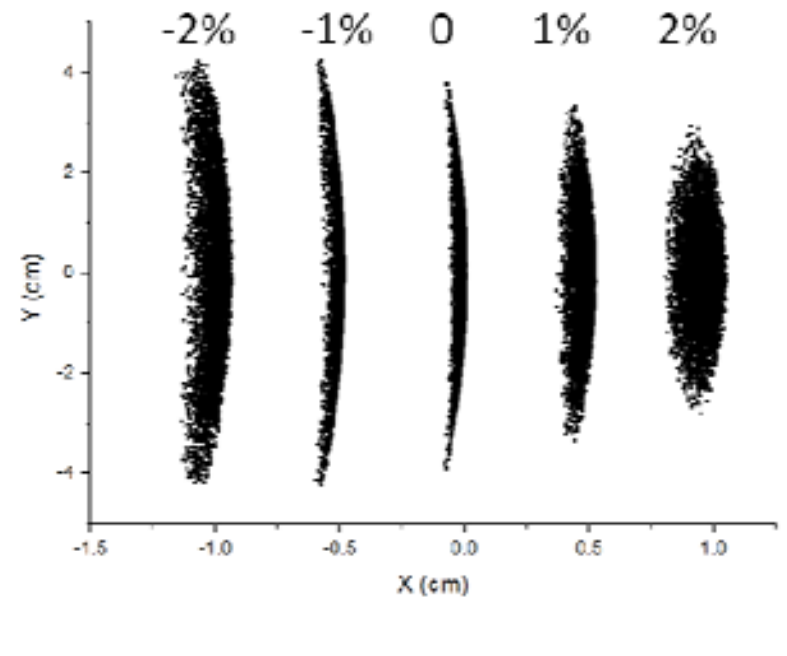

Fig.3 Simulation of analyzing ability of bending magnet

The energy of center beam is 15MeV. The other beams have ±1% and ±2% different energy compare with the center beam. The energy spread of every beam is ±0.000001%.

Back focus lens: A quadrupole-doublet lens is used to focus the proton beam to the end. The inner radius of lens is 50mm, the length of lens is 300mm and the distance is 20mm between each other. When the magnet fields of lens are -0.278, 0.368 kG/cm, the radius of beam focus is 9mm.

## 3 Efficiency of transport

From the simulation of 15MeV proton beam by program Track, we can get the efficiency of transport. The result is shown in Fig.4.

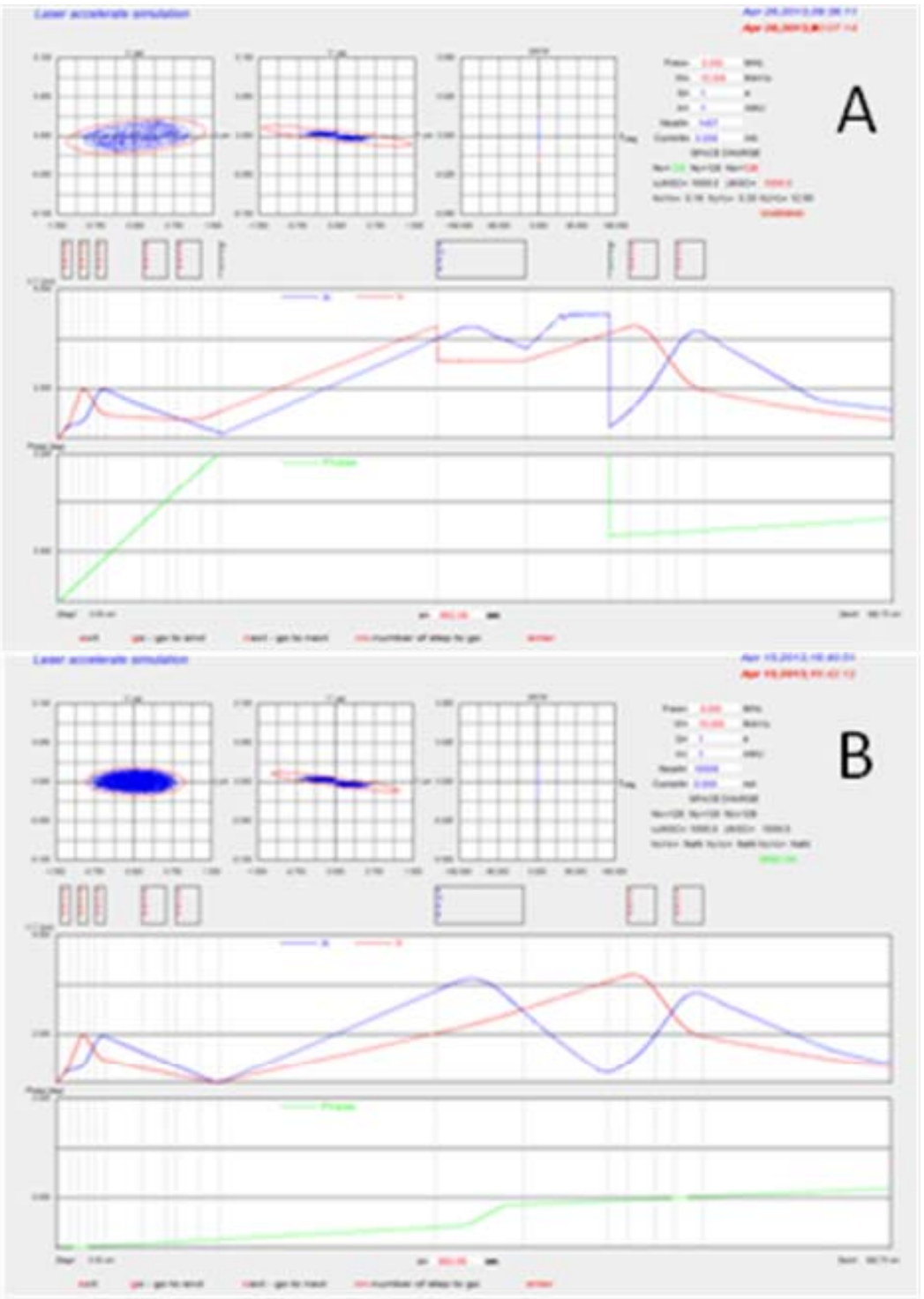

Fig.4 The efficiency simulation of 15MeV proton beam

A: The energy spread is ±15%, B: The energy spread is ±1%.

Most of protons with the energy spread of ±15% can be delivered to the bending magnet after being focused by collecting lens. Some protons with big energy spread impact the vacuum tube of deflection magnet in vertical direction, the others impact the vacuum tube and analysis aperture outside of bending magnet in horizontal direction. Finally, we can get the proton beam within the energy spread of ±1%.

The efficiency of all protons is nearly 14%. The efficiency of needed protons, which means the protons with initial energy spread smaller than ±1%, is bigger than 99%. Considering the magnet field distortion of quadrupole lens, the efficiency of needed protons is nearly 94%.

## 4 Transport of higher energy beam

The proton beam produced by the laser accelerator has wide energy spectrum. In order to widen the scope of application, the transport of higher energy proton beam is also taken into consideration. The transport of higher energy beam is achieved just by changing the magnet fields of lenses and bending magnet. The transport of 50MeV proton beam is shown in Fig.5.

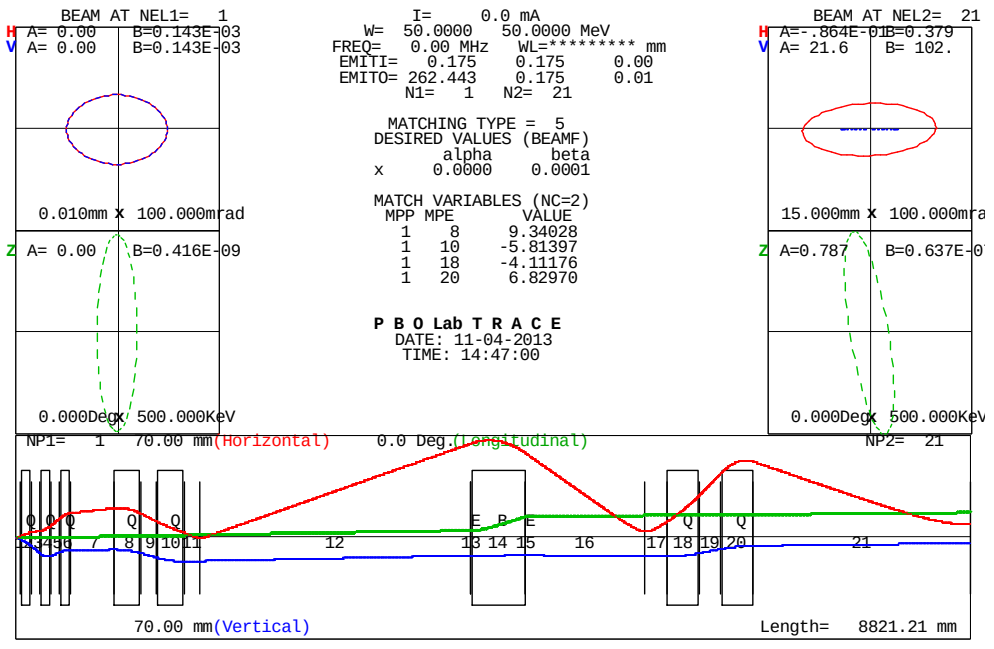

Fig.5 The transport of 50MeV proton beam

## 5 Summary

The beam line of CLAPA is preliminary designed by common transport elements. From the simulations of program Trace-3D and Track, the beam line is able to deliver the proton beam with energy of 1~50MeV and energy spread within ±1%. In future, the design will be improved if we get more accurate beam parameters.